\documentclass[aps,prd,a4paper,twocolumn,amsmath,showpacs,superscriptaddress,nofootinbib,preprintnumbers]{revtex4-1}

\usepackage{verbatim}
\usepackage[T1]{fontenc}
\usepackage[utf8]{inputenc}
\usepackage[american]{babel}
\usepackage{epsfig}
\usepackage{graphicx}
\usepackage{booktabs}
\usepackage{multirow}
\usepackage{dcolumn}
\usepackage{amsmath}
\usepackage{mathtools}
\usepackage{amsfonts}
\usepackage{amssymb}
\usepackage{epstopdf}
\usepackage{bm}
\usepackage{siunitx}
\usepackage{braket}
\usepackage{enumitem}
\usepackage{soul}
\usepackage[table]{xcolor}
\usepackage{color}
\usepackage{transparent}
\usepackage{pifont}

\definecolor{navyblue}{rgb}{0.0, 0.0, 0.5}
\definecolor{royalblue}{rgb}{0.25, 0.41, 0.88}
\definecolor{cadmiumgreen}{rgb}{0.0, 0.42, 0.24}
\definecolor{blue-violet}{rgb}{0.54, 0.17, 0.89}
\definecolor{darkviolet}{rgb}{0.58, 0.0, 0.83}
\definecolor{orange(colorwheel)}{rgb}{1.0, 0.5, 0.0}

\usepackage{hyperref}
\hypersetup{
    colorlinks=true, % false: boxed links; true: colored links
    linkcolor=royalblue, % color of internal links (change box color with linkbordercolor)
    citecolor=magenta}

\newcommand\ee{\end{equation}}
\newcommand\be{\begin{equation}}
\newcommand\eea{\end{eqnarray}}
\newcommand\bea{\begin{eqnarray}}

\newcommand{\alens}{A_{\rm lens}}

\usepackage{booktabs}
\usepackage{multirow}
\usepackage{dcolumn}
\usepackage{colortbl}

\definecolor{magenta(process)}{rgb}{1.0, 0.0, 0.56}

\definecolor{darkspringgreen}{rgb}{0.09, 0.45, 0.27}

\definecolor{royalblue(web)}{rgb}{0.25, 0.41, 0.88}
\begin{document}

\title{Constraining Dark Energy Dynamics in Extended Parameter Space}  

\author{Eleonora Di Valentino}
\email{eleonora.di_valentino@iap.fr}
\affiliation{Institut d'Astrophysique de Paris (UMR7095: CNRS \& UPMC- Sorbonne Universities), F-75014, Paris, France}
\affiliation{Sorbonne Universit\'es, Institut Lagrange de Paris (ILP), F-75014, Paris, France}
\author{Alessandro Melchiorri}
\email{alessandro.melchiorri@roma1.infn.it}
\affiliation{Physics Department and INFN, Universit\`a di Roma ``La Sapienza'', Ple Aldo Moro 2, 00185, Rome, Italy}
\author{Eric V. Linder}
\email{evlinder@lbl.gov}
\affiliation{Berkeley Center for Cosmological Physics \& Berkeley Lab, University of California, Berkeley, CA 94720, USA}
\affiliation{Energetic Cosmos Laboratory, Nazarbayev University, Astana, Kazakhstan 010000} 
\author{Joseph Silk}
\email{silk@iap.fr}
\affiliation{Institut d'Astrophysique de Paris (UMR7095: CNRS \& UPMC- Sorbonne Universities), F-75014, Paris, France}
\affiliation{AIM-Paris-Saclay, CEA/DSM/IRFU, CNRS, Univ. Paris VII, F-91191 Gif-sur-Yvette, France}
\affiliation{Department of Physics and Astronomy, The Johns Hopkins University Homewood Campus, Baltimore, MD 21218, USA}
\affiliation{BIPAC, Department of Physics, University of Oxford, Keble Road, Oxford OX1 3RH, UK}

\preprint{}
\begin{abstract}
Dynamical dark energy has been recently suggested as a promising and physical way to solve the 3 sigma tension on the value of the Hubble constant $H_0$ between the direct measurement of Riess et al. (2016)  (R16, hereafter) and the indirect constraint from Cosmic Microwave Anisotropies obtained by the Planck satellite under the assumption of a $\Lambda$CDM model.
In this paper, by parameterizing dark energy evolution using the $w_0$-$w_a$ approach, and considering a $12$ parameter extended scenario, we find that: a) the tension on the Hubble constant can indeed be solved with dynamical dark energy, b) a cosmological constant is ruled out at more than $95 \%$ c.l.\ by the Planck+R16 dataset, and  c) all of the standard quintessence and half of the "downward going" dark energy model space (characterized by an equation of state that decreases with time) is also excluded at more than $95 \%$ c.l. These results are further confirmed when cosmic shear, CMB lensing, or SN~Ia luminosity distance data are also included. 
However, tension remains with the BAO dataset. A cosmological constant and small portion of the freezing quintessence models are still in agreement with the Planck+R16+BAO dataset at between 68\% and 95\% c.l. Conversely, for Planck plus a phenomenological $H_0$ prior, both thawing and freezing quintessence models prefer a Hubble constant of less than 70 km/s/Mpc. The general conclusions hold also when considering models with non-zero spatial curvature.
\end{abstract}
\maketitle

\section{Introduction}

Recent measurements of the Cosmic Microwave Background (CMB) anisotropies made by the Planck satellite have provided strong confirmation of the $\Lambda$CDM model of structure formation based on cold dark matter (CDM), inflation and a cosmological constant $\Lambda$ (see e.g., \cite{planck2015} and the more recent analysis of \cite{plancknewtau}). However a few interesting tensions and anomalies are emerging that, albeit at low statistical significance, clearly justify the study of possible extensions to $\Lambda$CDM. 

While it is possible that tensions may arise from systematics in the measurements,  for the purposes of this article  we will take measurements from the CMB and other data sets at face value, and explore possible extension of the standard $\Lambda$CDM cosmology consistent with them. 

Tensions at the level of 95\% confidence seem present when the CMB temperature and polarization angular spectra $C_{\ell}$ are analyzed under $\Lambda$CDM. Indeed, the constraints on the $6$ parameters of the $\Lambda$CDM model are in disagreement at the $95\%$ c.l.\ when derived from data taken at small angular scales ($\ell>1000$) or large and intermediate angular scales ($\ell<1000$) (see discussion in \cite{Addison:2015wyg} and \cite{Aghanim:2015xee}).
Moreover, the value of the optical depth parameter $\tau=0.055\pm0.009$ recovered from Planck HFI large angular scale polarization measurements \cite{plancknewtau} is $1.7$ standard deviations lower than the constraint $\tau=0.099\pm0.024$ obtained using Planck TT + lowl (\cite{planck2015}), i.e. temperature in the full multipole range and polarization data at small angular scales ($\ell>29$).

This inadequacy of $\Lambda$CDM in providing a perfect fit to the Planck CMB temperature and polarization angular spectra is probably most evident in the anomalous value of the $A_{\rm lens}$ parameter that controls the amount of gravitational lensing in small-scale anisotropies (see e.g. \cite{calens} for a definition). Indeed, from the most recent analysis of Planck data \cite{plancknewtau}, one obtains the constraint $\alens=1.15^{+0.13}_{-0.12}$ at $95 \%$ c.l., i.e.\ higher than the value $\alens=1$ expected in $\Lambda$CDM at more than two standard deviations. 

While the $\alens$ and $\tau$ anomalies are {\it internal} inconsistencies present in current Planck data, the parameters derived from the Planck data, assuming $\Lambda$CDM, are also in tension with other {\it external}, i.e.\ non-CMB, datasets.

The current most statistically relevant tension is probably between the value of the Hubble constant derived from Planck and the one directly obtained from local luminosity distance measurements. Indeed, the recent value reported by Riess et al. (R16, hereafter)\ of $H_0=73.24 \pm 1.74$ km/s/Mpc at $68 \%$ c.l. ($R16$ hereafter, \cite{R16}) is in tension at more than $3$ standard deviations with the Planck result of $H_0=66.93\pm0.62$ km/s/Mpc at $68 \%$ c.l.\ obtained under the assumption of $\Lambda$CDM \cite{plancknewtau}. This tension is partially confirmed by the recent determinations from the H0LiCOW strong lensing survey that reports a value of the Hubble constant of $H_0=71.9_{-3.0}^{+2.4}$ km/s/Mpc at $68 \%$ c.l.\ \cite{holicow}, i.e.\ higher than the Planck value and more consistent with the $R16$ result, albeit at moderate statistical significance.

Another puzzling tension is the persisting discrepancy between the constraints on the $\sigma_8$ vs $\Omega_m$ plane obtained by Planck  and cosmic shear surveys such as CFHTLenS \cite{Heymans:2012gg} and KiDS-450 \cite{Hildebrandt:2016iqg} (again both under the assumption of  $\Lambda$CDM). Considering the parameter $S_8=\sigma_8 \sqrt{\Omega_m/0.3}$, the recent results from the KiDS-450 survey are in tension with the Planck results at about $2.3$ standard deviations. 
 
These various tensions have already motivated several studies of extensions of the $\Lambda$CDM scenario
(see e.g. \cite{dms,hodev,Zhao:2017urm,Yang:2017amu,Prilepina:2016rlq,Santos:2016sog,Kumar:2016zpg,joudaki,Karwal:2016vyq,Ko:2016uft,Bernal:2016gxb,Archidiacono:2016kkh,Qing-Guo:2016ykt,Zhang:2017idq,zhao,Sola:2017jbl,brust}). While one or two parameter extensions have been widely considered in the literature,  some of us recently took a more drastic approach of performing an analysis in a $12$ parameter space (\cite{dms}, \cite{hodev}), essentially doubling the degrees of freedom of $\Lambda$CDM. 

In principle, by increasing the number of parameters, one should solve any tension in the data \footnote{The famous sentence, attributed to John Von Neumann by Enrico Fermi: "With four parameters I can fit an elephant, and with five I can make him wiggle his trunk." should definitely be kept in mind here. In our defense, we quote Wolfgang Pauli's rejoinder to Von Neumann: "If proving something mathematically made a great physicist, you would be a great physicist." That is, we let the observational data determine which physical parameters have their concordance model values.}. However there are in our opinion several reasons that can motivate this kind of analysis. 

First of all, the $\Lambda$CDM model is clearly at risk of presenting an oversimplification of the physics that drives the evolution of our universe. There is indeed no reason to fix the sum of neutrino masses to $\Sigma m_{\nu} =0.06\,$eV, i.e.\ to the minimum value admitted by current oscillation experiments, or to believe that the mysterious dark energy component that produces the current acceleration of the universe can be completely parametrized by just a constant energy density term. 

Secondly, an indication for a persistent anomaly in $12$ parameter space should be considered as more robust with respect to a similar anomaly with similar statistical confidence but obtained in a much more reduced parameter space. In other words, parameter constraints in extended scenarios should be regarded as more conservative with respect to those obtained under $\Lambda$CDM.

Finally, some of the tensions can be solved at the same time by introducing two, sometimes very different, extensions. For example, the tension of the Hubble parameter can be solved either by increasing the effective number of neutrino species at recombination $N_{\rm eff}$ or by considering a dark energy equation of state $w<-1$ (see discussion in \cite{R16}). By varying both parameters simultaneously, one can understand which of the two extensions can better solve this tension and the interaction between them. Indeed, we will find that there is no preference for extra neutrino species ("dark radiation") when allowing for dynamical dark energy. 

Recently, in \cite{hodev}, it has been shown that if one considers a $12$ parameter extended scenario, the tension on the Hubble parameter can be solved by a dark energy equation of state $w<-1$, while the neutrino effective number is fully compatible with the standard expected value of $N_{\rm eff}=3.046$. The tension between the Planck result and the R16 value can therefore be better explained by introducing a dynamical form of dark energy, with a cosmological constant excluded at more than $95 \%$ c.l.
A similar conclusion has been recently obtained in \cite{zhao} in a complementary approach. 

Since a dark energy component with a constant-with-redshift equation of state suffers from the usual "why now?" problem of the cosmological constant and is generally not expected in physically-motivated scenarios, inferring $w\neq-1$ from the data immediately triggers interest in a possible evolution of the dark energy equation-of-state with redshift.
This is the focus of our analysis here: constraining the evolution of dark energy in an extended parameter space and finding the preferred model or region, if any, in a combined Planck+R16 analysis.

One of the simplest physical alternatives to a cosmological constant is a dynamical scalar field (see e.g.\ \cite{quintessence1},\cite{quintessence2}, \cite{quintessence3}). We treat this in terms of a phenomenological equation of state. Indeed, since we will allow equation-of-state behavior that crosses $w=-1$, this must involve an effective scalar field, e.g.\ the sum of two minimally coupled scalar fields. We keep the dark energy sound speed fixed at the speed of light, and include perturbations through the standard PPF module of CAMB \cite{whu}. Quintessence (minimally coupled, canonical scalar fields) in our universe can be divided into two types:
"thawing" models (\cite{thawing1},\cite{thawing2},\cite{thawing3}), where $w(a)$ is a growing function of $a$ from cosmological constant behavior in the early universe, and "freezing" models (\cite{freezing1},\cite{freezing2},\cite{freezing3}) where, on
the contrary, the equation-of-state is a decreasing function of the scale factor $a$, approaching a cosmological constant. Since the behaviors we consider include the unquintessential ability to cross $w=-1$, we instead refer to "upwards going" ($w$ increasing with $a$) and "downwards going" ($w$ becoming more negative with $a$) classes. 

For the standard parameterization 
\begin{equation}
w(a)=w_0+(1-a)w_a
\end{equation} 
where $w_0$ is the present value of the equation of state, and $w_a$ a measure of the equation-of-state time variation, "upwards-going" models correspond to $w_a<0$, and "downwards-going" models to $w_a>0$. The sign of $w_a$ can give an idea of the type of dark energy evolution, and the sign of $1+w_0$ determines whether the dark energy is currently phantom ($w_0<-1$) or not. These two constants are to be determined by the observations.

\section{Method}

The goal of this paper is to constrain dark energy dynamics in a considerably extended parameter space. For our theoretical baseline, we consider $12$ parameters that are varied simultaneously in a range of external, conservative, priors listed in Table \ref{priors}. These are the six parameters of the standard $\Lambda$CDM model, i.e.\ the Hubble constant $H_0$, the baryon energy density $\Omega_bh^2$, the 
cold-dark-matter energy density $\Omega_ch^2$, the amplitude and spectral index of the primordial scalar perturbations $A_s$ and $n_s$ (at pivot scale $k_0=0.05 h\,{\rm Mpc}^{-1}$), and the reionization optical depth $\tau$. Moreover, we add variations in $5$ more parameters, i.e.\ the total neutrino mass for the $3$ standard neutrinos $\Sigma m_\nu$, the two dark energy equation-of-state parameters $w_0$ and $w_a$, the running of the scalar spectral index $dn_s/d\ln k$, and the effective number of relativistic degrees of freedom $N_{\rm eff}$. Finally, we also consider variation in the gravitational lensing
amplitude of the CMB angular spectra $\alens$. This scales the CMB lensing strength on all scales, relative to the prediction of the model being considered. 

We also consider two more scenarios in addition to our baseline model. In one case, we fix $\alens=1$ since, at the moment, the origin of this anomaly is still unclear. In practice, $\alens$ is an effective parameter that could just be compensating for a statistical fluke in the data. It is therefore important to investigate if the inclusion of $\alens$ has an impact on the constraints on the dark energy equation of state.
As a second additional scenario, we also fix  $\alens=1$ but we now vary the curvature density $\Omega_k$ since, again, the Planck angular spectrum data within the $\Lambda$CDM model suggest a closed universe at about two standard deviations.

We analyze these cosmological parameters by using, firstly, the temperature and polarization CMB angular power spectra released by Planck 2015 \cite{Aghanim:2015xee}. This dataset includes the large angular-scale temperature and polarization anisotropy measured by the Planck LFI experiment and the small-scale anisotropies measured by Planck HFI, and we refer to it as "Planck". 
We then consider the following additional data sets:

\begin{itemize}

\item{\bf R16}:
As "R16", we consider an external gaussian prior on the Hubble constant $H_0=73.24\pm1.74$ km/s/Mpc at $68 \%$ c.l., as measured by \cite{R16}.

\item{\bf BAO}: We make use of the baryon acoustic oscillation data from 6dFGS \cite{beutler2011}, SDSS-MGS \cite{ross2014}, BOSSLOWZ \cite{anderson2014} and CMASS-DR11 \cite{anderson2014} surveys as in \cite{planck2015}. We refer to this dataset as "BAO".

\item{\bf JLA}:
We include luminosity distances of supernovae Type Ia from the "Joint Light-curve Analysis" derived from the SNLS and SDSS catalogs \cite{JLA}. We refer to this dataset as "JLA".

\item{\bf WL}:
We consider the weak lensing galaxy data from the CFHTlens \cite{Heymans:2012gg} survey with the priors and conservative cuts to the data as described in \cite{planck2015}.

\item{\bf Lensing}:
We indicate with "lensing" the information we can derive from CMB lensing from the Planck trispectrum detection \cite{Ade:2015zua}.

\end{itemize} 

We also study the effect on the preferred type of dark energy behavior from variations in the Hubble constant prior, from values of 66 to 74 km/s/Mpc. This can be viewed as a phenomenological study, or values from different data sets or analyses (e.g.\ \cite{rigault2015}). 

In order to derive constraints on the parameters, we use the July 2015 version of the publicly available Monte Carlo Markov Chain package \texttt{cosmomc} \cite{Lewis:2002ah}, that has a convergence diagnostic based on the Gelman and Rubin statistic and includes the support for the Planck data release 2015 Likelihood Code \cite{Aghanim:2015xee} (see \url{http://cosmologist.info/cosmomc/}), implementing an efficient sampling by using the fast/slow parameter decorrelations \cite{Lewis:2013hha}.
While we focus our attention on cosmological parameters, we also vary the foreground parameters as described in \cite{Aghanim:2015xee} and \cite{planck2015}.

\begin{table}
\begin{center}
\begin{tabular}{c|c}
% \hline
Parameter                    & Prior\\
\hline
$\Omega_{\rm b} h^2$         & $[0.005,0.1]$\\
$\Omega_{\rm c} h^2$       & $[0.001,0.99]$\\
$\tau$                       & $[0.01,0.8]$\\
$n_s$                        & $[0.8, 1.2]$\\
$\log[10^{10}A_{s}]$         & $[2,4]$\\
$\Theta_{\rm s}$             & $[0.5,10]$\\ 
$\sum m_\nu$ (eV)               & $[0,5]$\\
$w_0$ & [-3,0.3]\\
$w_a$ & [-2,2]\\
$N_{\rm eff}$ & [0.05,10]\\ 
$\frac{dn_s}{d\ln k}$ & [-1,1]\\
$\alens$ & [0,10]\\
$\Omega_k$ & [-0.3,0.3] \\
% \hline
\end{tabular}
\end{center}
\caption{External flat priors on the cosmological parameters assumed in this paper. Note that $\Theta_{\rm s}$ is used for the likelihood evaluation but $H_0$ is quoted for parameter constraints.}
\label{priors}
\end{table}

\section{Results}

\subsection{12 Parameter analysis}

%\onecolumngrid
\begin{figure*}
\centering
\begin{tabular}{cc}
\includegraphics[width=0.95\columnwidth]{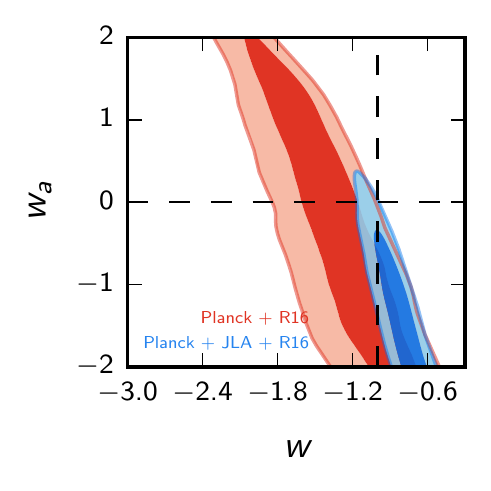} & \includegraphics[width=0.95\columnwidth]{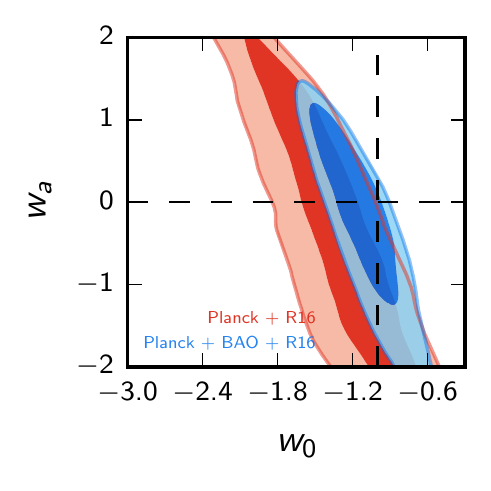}  \\
\includegraphics[width=0.95\columnwidth]{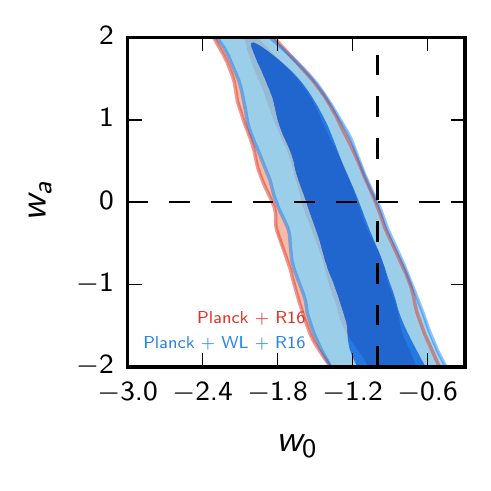}  &
\includegraphics[width=0.95\columnwidth]{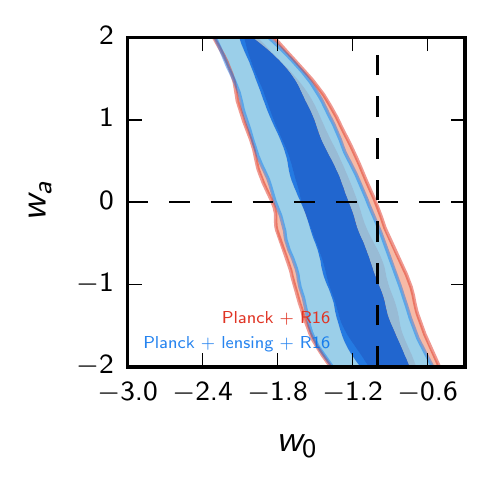} \\
\end{tabular}
\caption{68.3\% and 95.4\% confidence level constraints on the $w_0$--$w_a$ plane in a $12$ parameter extended space for the Planck+R16 data, and combined with several other datasets. Only in the case
of Planck+R16+BAO (top right panel), is a cosmological constant within the 95.4\% c.l. In all other cases a cosmological
constant and the region ($w_0>-1$, $w_a>0$) is excluded at more than $95.4\%$ c.l.}
\label{fig1}
\end{figure*}
%\twocolumngrid

\begin{table*}
\begin{center}\footnotesize
\scalebox{0.93}{\begin{tabular}{lcccccccc}
\hline \hline
         & Planck & Planck  & Planck & Planck & Planck &Planck&Planck\\                     
         &      &        +R16  & +BAO & +R16+BAO      &    +R16+JLA & +R16+WL&+R16+lensing\\  
\hline
\hspace{1mm}\\

$\Omega_{\textrm{b}}h^2$& $0.02223\,\pm 0.00028 $& $0.02223\,\pm 0.00028$    & $0.02238\,\pm0.00027 $&$0.02251\,\pm0.00027 $& $0.02251\,\pm 0.00024$& $0.02236\,\pm 0.00027$ & $0.02205\,\pm 0.00027$   \\
\hspace{1mm}\\

$\Omega_{\textrm{c}}h^2$& $0.1186\,\pm 0.0035$& $0.1186\,\pm 0.0034$    & $0.1185\,\pm0.0034 $&$0.1210\,\pm0.0035 $& $0.1203\,\pm 0.0033$ & $0.1189\,\pm0.0035$ & $0.1180\,\pm 0.0035$   \\
\hspace{1mm}\\

$\tau$& $0.059\,\pm 0.021$& $0.058\,\pm0.021$    & $0.059\,\pm 0.021$&$0.059\,\pm 0.021$& $0.058\,\pm 0.022$& $0.051\,\pm 0.020$ & $0.059\,\pm 0.021$    \\
\hspace{1mm}\\

$n_s$& $0.963\,\pm 0.013$& $0.963\,\pm 0.012$    & $0.967\,\pm 0.012 $& $0.975\,\pm 0.012 $& $0.974\,\pm 0.011$& $0.969\,\pm 0.012$ & $0.957\,\pm 0.013$   \\
\hspace{1mm}\\

$\log(10^{10}A_S)$& $3.048\,\pm 0.044$& $3.047\,\pm0.044$    & $3.048\,\pm 0.044$& $3.056\,\pm 0.044$& $3.052\,\pm 0.045$& $3.032\,^{+0.040}_{-0.046}$ & $3.045\,\pm 0.044$    \\
\hspace{1mm}\\

$H_0$ &      $77\,^{+20}_{-10}$&      $ 73.9\,\pm 2.0$ & $ 65.6\,^{+2.1}_{-3.1}$   &$ 71.3\,\pm 1.9$   &  $ 71.3\,\pm 1.5$ &  $ 73.9\,\pm 2.0$  &  $ 74.0\,\pm 2.0$ \\
\hspace{1mm}\\

$\sigma_8$   & $ 0.81\,_{-0.12}^{+0.14}$   & $ 0.799\,\pm0.053$   & $ 0.765\,\pm 0.036$ & $ 0.796\,\pm 0.040$ &  $ 0.810\,_{-0.034}^{+0.049}$ &  $ 0.777\,_{-0.051}^{+0.056}$ &  $ 0.814\,\pm0.045$  \\
\hspace{1mm}\\

$\sum m_{\nu}$ [eV] &      $0.52\, _{-0.45}^{+0.20}$&      $0.51\, _{-0.45}^{+0.20}$ & $<0.557$& $0.34\, _{-0.27}^{+0.15}$   & $<0.648$ & $0.57\, _{-0.42}^{+0.26}$ & $0.46\, _{-0.27}^{+0.23}$  \\
\hspace{1mm}\\

$w_0$ &  $-1.46\,\pm0.59$ &  $-1.39\,_{-0.32}^{+0.39}$  & $-0.71\,_{-0.16}^{+0.29}$&$-1.14\,\pm0.21$&  $-0.86\,_{-0.10}^{+0.15}$ &  $-1.35\,_{-0.33}^{+0.38}$ &  $-1.42\,_{-0.32}^{+0.37}$   \\
\hspace{1mm}\\

$w_a$ &  $-0.2\,_{-1.6}^{+0.8}$ &  $-0.2\,_{-1.6}^{+0.8}$  & $<0.179$& $-0.16\,_{-0.64}^{+0.97}$&  $<-0.134$ &  unconstrained &  unconstrained   \\
\hspace{1mm}\\

$N_{\rm eff}$ &  $3.01\,\pm0.25$ &  $3.01\,\pm0.25$  & $3.05\,\pm0.25$& $3.23\,\pm0.25$&  $3.20\,\pm0.22$ &  $3.11\, \pm 0.25$ &  $2.92\,\pm0.25$  \\
\hspace{1mm}\\

$\frac{dn_s}{d\ln k}$ &  $-0.0004\,\pm 0.0089$ &  $-0.0003\,\pm0.0088$  & $-0.0015\,\pm 0.0084$& $0.0023\,\pm 0.0082$&  $0.0021\,\pm0.0083$ &  $0.0030\,\pm0.0085$ &  $-0.0020\,\pm0.0087$   \\
\hspace{1mm}\\

$\alens$ &  $1.21\,_{-0.13}^{+0.10}$ &  $1.20\,_{-0.11}^{+0.10}$  & $1.18\,^{+0.09}_{-0.11}$& $1.21\,\pm 0.10$&  $1.18\,_{-0.11}^{+0.09}$ &  $1.25\, _{-0.11}^{+0.10}$ &  $1.061\pm0.075$   \\
\hspace{1mm}\\

\hline
\hline

\end{tabular}}
\caption{$68 \%$ c.l.\ constraints on cosmological parameters in our extended $12$ parameter scenario from different combinations of datasets. If only upper limits are shown, they are at 95\% c.l. Note that $\sigma_8$ is a derived parameter.} 
\label{table1}
\end{center}
\end{table*}

The main results of our analysis are reported in Table \ref{table1} where we report the constraints at $68 \%$ c.l.\ on the $12$ parameters of our theoretical framework, using different combinations of datasets.

Since the goal of this paper is to constrain dynamical dark energy in this extended parameter space, we show the  confidence levels contour plots in the $w_0$-$w_a$ plane in Figure \ref{fig1} for the Planck+R16 case as well as in combination with several other datasets. 

The R16 Hubble constant prior has the main effect of significantly ruling out at more than $95 \%$ c.l.\ the region $w_0 \ge -1$ and $w_a \ge 0$. In other words, the R16 prior not only rules out a cosmological constant as already discussed in \cite{hodev}, but all the freezing quintessence models and the half of the downward going dark energy models with $w_a>0$ and $w_0>-1$.
As one can clearly see from Figure~\ref{fig1}, once the R16 prior is included, Planck data combined with the JLA, lensing, or WL dataset all
disfavor the $(w_a\ge0,w_0\ge-1)$ region (though the last two do not significantly add to the constraints currently). BAO data are however still in tension with the Planck+R16 dataset.  The tension between BAO and Planck+R16 (already noticed in \cite{hodev} and \cite{Bernal:2016gxb}) is present also in our $12$ parameters analysis
and can be clearly seen in the two different constraints for the Hubble constant: $H_0=73.9\pm2.0$ km/s/Mpc for Planck+R16 and $H_0=65.6^{+3.1}_{-2.1}$ km/s/Mpc at $68 \%$ c.l.  for Planck+BAO, therefore inconsistent at about $2.9$ standard deviations. When the BAO dataset is included, a small region with $w_a\ge0$, $w_0\ge-1$ (including the cosmological constant) is back in agreement with data in between 68\% and 95\% c.l..

Moreover, it is interesting to examine the case of the other constraining external data set: the supernovae. The combined Planck+R16+JLA dataset is shown in the top left panel of Figure~\ref{fig1}. From this dataset almost the entire $w_a>0$ region seems to be disfavored, i.e.\ the whole class of downward going models is excluded at more than $95 \%$ c.l.\ from this combination of data. We also note that of the allowed region for upward going models, almost none of it corresponds to standard thawing quintessence as we discuss later, i.e.\ both of the standard quintessence classes are disfavored. 

Looking at the other parameter constraints in Table~\ref{table1},  we also notice that the $\alens$ parameter is always larger than the expected value of $\alens=1$ for any combination of data, with the  exception of the case when the CMB lensing dataset is included. However, the conclusions on dark energy seem unaffected by this. Considering the contour plots in Figure~\ref{fig1}, bottom right panel, we indeed see that the exclusion of the region ($w_a\ge0$, $w_0\ge-1$) is stable also for the Planck+R16+lensing dataset. This is somewhat reassuring. However, the Planck+R16+lensing dataset also suggests the presence of a neutrino mass at $\Sigma m_{\nu} \approx 0.46$ eV at slightly below two standard deviations. 

From Planck+R16 we found $S_8=0.754\pm0.041$ that is in complete agreement with the value $S_8=0.745\pm0.039$ from the KIDS-450 lensing survey \cite{joudaki}.

\begin{figure}
\centering
\includegraphics[width=0.95\columnwidth]{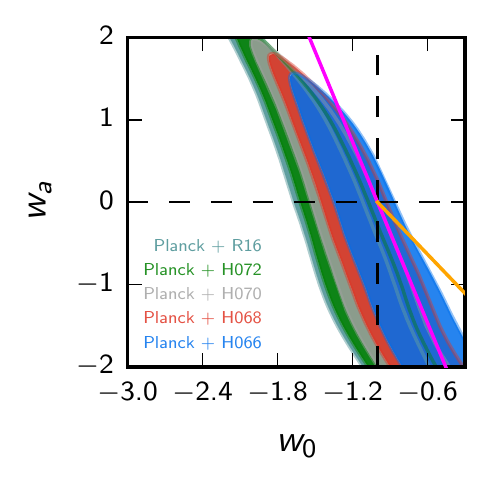} 
\caption{The effects of shifting the $H_0$ prior are illustrated for the Planck+$H_0$ prior set in the 12 parameter extended space. $68\%$ c.l.\ constraints on the $w_0$ vs $w_a$ plane are shown for five different $H_0$ priors.The steep magenta line shows the "mirage" line giving the main geometric CMB degeneracy. The shallower orange half line shows the $w_0$-$w_a$ relation that many thawing quintessence models follow.}
\label{figH0}
\end{figure}

To highlight the importance of the Hubble constant prior, we show its effect in Figure~\ref{figH0}. Here we plot the Planck+R16 contour as in Figure~\ref{fig1}, but also indicate how the best fit $w_0$-$w_a$ values shift if we replace the R16 prior on $H_0$ with different central values (and the same absolute uncertainty). 
In addition, we overlay two model lines: one is the "mirage" line, where a time-varying dark energy gives the same distance to the CMB last scattering surface as for a $\Lambda$CDM model, and the other line is centered on the typical behavior of thawing quintessence. These lines are respectively $w_a=-3.66(1+w_0)$ \cite{0708.0024} and $w_a=-1.58(1+w_0)$ \cite{1501.01634}. We see that as $H_0$ decreases from R16, the data become more consistent with the cosmological constant, as well as with the thawing behavior, and the contour also edges into the upper right quadrant where freezing quintessence resides. When $H_0<70\,$km/s/Mpc then $\Lambda$ enters the 68\% CL. 

Note that the contours remain roughly parallel to the mirage line as $H_0$ changes. This line reflects preservation of the distance to CMB last scattering $d_{\rm lss}$, for a fixed $\Omega_m$. One can get a rough estimate of the size of the shift of the contours when $H_0$ changes by looking at the ratio of derivatives $\partial d_{\rm lss}/\partial\Omega_m$ and $\partial d_{\rm lss}/\partial w$ for constant $w$, and translating to a shift in $H_0$ by assuming the physical matter density $\Omega_m h^2$ (well determined by the CMB) is also preserved. This gives $\Delta w\approx -3.3\Delta h$. This roughly gives the horizontal shift of the contours along the $w_a=0$ axis; in fact, the shift is somewhat greater because of the covariance between $w_0$ and $w_a$. 

We also show the thawing line, in the vicinity of which most thawing quintessence models lie (see, e.g., Figure 10 of \cite{0808.0189}). Note that this does not enter the 68\% CL until $H_0<70\,$km/s/Mpc. It is extremely difficult for a standard quintessence model to lie in the "superthawing" region between the thawing line and the $w_0=-1$ axis; recall that the thawing dynamics is driven by evolution during the matter-dominated epoch so superthawing would require violation of the  matter-dominated era or some extra impetus to the evolution, such as a second, phantom, field.

\subsection{$11$ Parameter analysis (fixing $\alens=1$)}

%\onecolumngrid
\begin{figure*}
\centering
\begin{tabular}{cc}
\includegraphics[width=0.95\columnwidth]{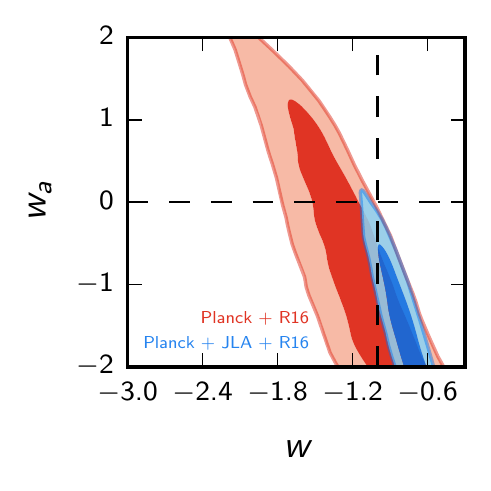} & \includegraphics[width=0.95\columnwidth]{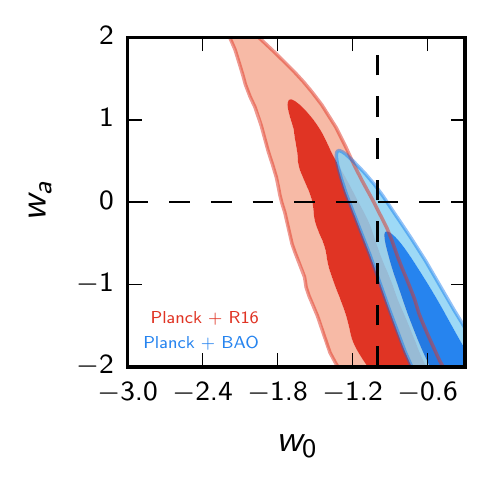}  \\
\includegraphics[width=0.95\columnwidth]{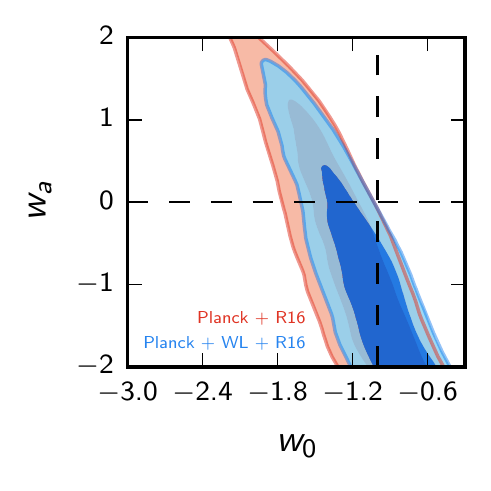}  &
\includegraphics[width=0.95\columnwidth]{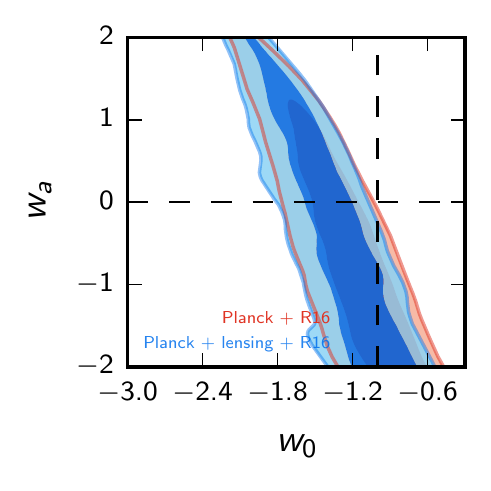} \\
\end{tabular}
\caption{68.3\% and 95.4\% constraints on the $w_0$--$w_a$ plane in an $11$ parameter extended space,  fixing $\alens=1$. Only in the case of Planck+BAO (top right panel) is a cosmological constant still within 95.4\% c.l. The allowed parameter space in the  ($w_0>-1$, $w_a>0$) region is also reduced relative to the $12$-parameter case. For Planck+R16+JLA, the entire region with $w_a>0$ is now even more strongly excluded.} 
\label{fig2}
\end{figure*}

%\twocolumngrid

Since $\alens>1$ for most of the data set combination, it is useful to further test the impact of a variation in the effective parameter $\alens$ on our results. We have performed an analysis in a restricted parameter space, fixing $\alens=1$, and report the results in 
Table~\ref{table2} and in Figure~\ref{fig2}.

\begin{table*}
\begin{center}\footnotesize
\scalebox{0.93}{\begin{tabular}{lccccccc}
\hline \hline
         & Planck & Planck  & Planck & Planck &Planck&Planck\\                     
         &      &        +R16  & +BAO       &    +R16+JLA & +R16+WL& +R16+lensing\\  
\hline
\hspace{1mm}\\

$\Omega_{\textrm{b}}h^2$& $0.02207\,\pm 0.00026 $& $0.02205\,\pm 0.00026$    & $0.02213\,\pm0.00023 $& $0.02231\,\pm 0.00021$& $0.02217\,\pm0.00025$ & $0.02197\,\pm 0.00025$   \\
\hspace{1mm}\\

$\Omega_{\textrm{c}}h^2$& $0.1175\,\pm 0.0033$& $0.1174\,\pm 0.0033$    & $0.1176\,\pm0.0033 $& $0.1198\,\pm 0.0031$ & $0.1171\,\pm0.0032$ & $0.1173\,\pm 0.0033$   \\
\hspace{1mm}\\

$\tau$& $0.078\,\pm 0.019$& $0.079\,\pm0.020$    & $0.080\,\pm 0.019$& $0.081\,\pm 0.019$& $0.075\,\pm0.018$ & $0.068\,^{+0.020}_{-0.017}$    \\
\hspace{1mm}\\

$n_S$& $0.954\,\pm 0.012$& $0.953\,\pm 0.012$    & $0.956\,\pm 0.010 $& $0.9650\,\pm 0.0095$& $0.958\,\pm0.010$ & $0.953\,\pm 0.011$   \\
\hspace{1mm}\\

$\log(10^{10}A_S)$& $3.086\,\pm 0.040$& $3.087\,\pm0.041$    & $3.091\,\pm 0.039$& $3.098\,\pm 0.039$& $3.078\,\pm0.037$ & $3.064\,\pm 0.038$    \\
\hspace{1mm}\\

$H_0$ &      $>62.4$&      $ 73.9\,\pm 2.0$ & $ 65.0\,^{+2.0}_{-2.9}$   &  $ 71.2\,\pm 1.5$ &  $ 73.7\,\pm 2.0$  &  $ 74.0\,\pm 2.0$ \\
\hspace{1mm}\\

$\sigma_8$   & $ 0.94\,_{-0.07}^{+0.13}$   & $ 0.873\,^{+0.037}_{-0.028}$   & $ 0.809\,^{+0.025}_{-0.029}$ &  $ 0.865\,_{-0.021}^{+0.026}$ &  $ 0.868\,^{+0.032}_{-0.024}$ &  $ 0.843\,\pm0.026$  \\
\hspace{1mm}\\

$\sum m_{\nu}$ [eV] &      $<0.608$&      $<0.621$ & $<0.332$   & $<0.306$ & $<0.501$ & $0.35\, _{-0.23}^{+0.17}$  \\
\hspace{1mm}\\

$w_0$ &  $-1.56\,^{+0.45}_{-0.61}$ &  $-1.27\,_{-0.25}^{+0.38}$  & $-0.70\,_{-0.15}^{+0.29}$&  $-0.84\,_{-0.09}^{+0.14}$ &  $-1.12\,_{-0.20}^{+0.33}$ &  $-1.37\,_{-0.29}^{+0.38}$   \\
\hspace{1mm}\\

$w_a$ &  $unconstrained$ &  $<1.30$  & $<0.055$&  $<-0.338$ &  $<0.889$ &  $unconstrained$   \\
\hspace{1mm}\\

$N_{\rm eff}$ &  $2.85\,\pm0.23$ &  $2.84\,\pm0.23$  & $2.88\,\pm0.22$&  $3.06\,\pm0.19$ &  $2.89\,\pm0.21$ &  $2.83\,\pm0.23$  \\
\hspace{1mm}\\

$\frac{dn_s}{d\ln k}$ &  $-0.0064\,\pm 0.0080$ &  $-0.0072\,\pm0.0077$  & $-0.0077\,\pm 0.0079$&  $-0.0040\,\pm0.0076$ &  $-0.0048\,^{+0.0071}_{-0.0082}$ &  $-0.0052\,\pm0.0076$   \\
\hspace{1mm}\\

\hline
\hline

\end{tabular}}
\caption{$68 \%$ c.l.\ constraints on cosmological parameters in our extended $11$-parameter scenario from different combinations of datasets. The $\alens$ parameter is kept fixed to $1$ in this analysis. If only upper/lower limits are shown, they are at 95\% c.l. Note that $\sigma_8$ is a derived parameter.}
\label{table2}
\end{center}
\end{table*}

The main conclusions about $w(a)$ obtained when varying $\alens$ remain robust for the case of $\alens$ fixed to the standard value of $1$.
As seen in Figure~\ref{fig2}, there is no significant shift in the best fit with respect to the $12$-parameter case reported in Figure~\ref{fig1}, while some of the allowed regions shrink moderately around it. For example, the region ($w_0<-1$, $w_a> 0$) is now even more excluded by the Planck+R16+JLA data set.

We however found that  the  BAO dataset is in this case even more in tension with Planck+R16 and that convergence in the MCMC chains is difficult to reach for the Planck+R16+BAO combined data. This is also clearly shown in Table~\ref{table2} where the constraints for the Hubble constant are now in tension by more than $3$ standard deviations. 
Further tension is present in the not complete overlap of the constraints in the $w_0-w_a$ plane for Planck+BAO and Planck+R16
(see Figure~\ref{fig2}, top right panel).
In this case we therefore decide not to include the Planck+BAO+R16
constraint.  It is however interesting to note that also  the Planck+BAO dataset  does not prefer the  ($w_0>-1$, $w_a> 0$) region.

We conclude that the effect of including $\alens$ as a twelfth fit parameter in the analysis is just to make the constraints in the ($w_0$, $w_a$) plane slightly narrower, with no significant shift in the best fit values.

As seen in Table~\ref{table2}, fixing $\alens=1$ results also in stronger bounds on neutrino masses, higher values for the r.m.s.\ mass fluctuation amplitude $\sigma_8$, and lower values for the effective neutrino number $N_{\rm eff} \sim 2.9$.

\subsection{$12$ Parameter analysis varying $\Omega_k$ instead of $\alens$}

We have also performed an analysis in $12$-parameter space by considering variation in curvature $\Omega_k$ instead of $\alens$. A value of $\Omega_k<0$ (closed universe) is slightly preferred by the Planck data set in the standard restricted (6+1) parameter space analysis, and it is therefore of interest to investigate this possibility in an extended parameter space.

\begin{table*}
\begin{center}\footnotesize
\scalebox{0.93}{\begin{tabular}{lccccccc}
\hline \hline
         & Planck & Planck  & Planck & Planck &Planck&Planck\\                     
         &      &     +lensing  & +BAO       &  +JLA & +R16+WL& +R16+lensing\\  
\hline
\hspace{1mm}\\

$\Omega_{\textrm{b}}h^2$& $0.02231\,\pm 0.00028 $& $0.02203\,\pm 0.00026$    & $0.02219\,\pm0.00027 $& $0.02231\,\pm 0.00028$& $0.02216\,\pm 0.00026$ & $0.02204\,\pm 0.00026$   \\
\hspace{1mm}\\

$\Omega_{\textrm{c}}h^2$& $0.1197\,\pm 0.0035$& $0.1181\,\pm 0.0035$    & $0.1178\,\pm0.0034 $& $0.1187\,\pm 0.0035$ & $0.1175\,\pm0.0032$ & $0.1180\,\pm 0.0034$   \\
\hspace{1mm}\\

$\tau$& $0.054\,^{+0.020}_{-0.024}$& $0.057\,\pm 0.021$    & $0.080\,\pm 0.019$& $0.063\,\pm 0.021$& $0.074\,\pm 0.020$ & $0.059\,\pm 0.021$    \\
\hspace{1mm}\\

$n_s$& $0.968\,\pm 0.012$& $0.957\,\pm 0.012$    & $0.958\,\pm 0.013 $& $0.965\,\pm 0.013$& $0.959\,\pm 0.012$ & $0.957\,\pm 0.012$   \\
\hspace{1mm}\\

$\log(10^{10}A_S)$& $3.039\,^{+0.041}_{-0.050}$& $3.042\,\pm 0.043$    & $3.090\,\pm 0.040$& $3.058\,\pm 0.044$& $3.078\,^{+0.043}_{-0.039}$ & $3.045\,\pm 0.043$    \\
\hspace{1mm}\\

$H_0$ &      $54\,^{+7}_{-20}$&      $ 69\,^{+10}_{-20}$ & $ 65.1\,^{+2.2}_{-3.0}$   &  $ 61.1\,^{+3.5}_{-4.1}$ &  $ 73.7\,\pm 2.0$  &  $ 74.3\,\pm 2.1$ \\
\hspace{1mm}\\

$\sigma_8$   & $ 0.74\,_{-0.16}^{+0.09}$   & $ 0.80\,^{+0.11}_{-0.15}$   & $ 0.811\,\pm 0.034$ &  $ 0.807\,_{-0.034}^{+0.046}$ &  $ 0.866\,\pm0.034$ &  $ 0.847\,\pm0.026$  \\
\hspace{1mm}\\

$\sum m_{\nu}$ [eV] &      $0.55\, _{-0.40}^{+0.25}$&      $0.47\, \pm 0.23$ & $<0.342$   & $<0.630$ & $<0.502$ & $0.43\, \pm 0.22$  \\
\hspace{1mm}\\

$w_0$ &  $unconstrained$ &  $-1.44\,_{-0.58}^{+0.85}$  & $-0.69\,_{-0.16}^{+0.27}$&  $-1.11\,_{-0.17}^{+0.25}$ &  $-1.12\,_{-0.24}^{+0.41}$ &  $-1.69\,_{-0.45}^{+0.58}$   \\
\hspace{1mm}\\

$w_a$ &  $unconstrained$ &  $-0.2\,_{-1.6}^{+0.7}$  & $<0.011$&  $<0.617$ &  $<1.03$ &  $unconstrained$   \\
\hspace{1mm}\\

$N_{\rm eff}$ &  $3.11\,\pm0.25$ &  $2.92\,\pm0.24$  & $2.91\,\pm0.24$&  $3.03\,\pm0.25$ &  $2.91\,\pm 0.23$ &  $2.91\,\pm0.24$  \\
\hspace{1mm}\\

$\frac{dn_s}{d\ln k}$ &  $0.0038\,\pm 0.0089$ &  $-0.0012\,\pm0.0087$  & $-0.0067\,\pm 0.0087$&  $0.0000\,\pm0.0088$ &  $-0.0043\,_{-0.0076}^{+0.0092}$ &  $-0.0013\,\pm0.0085$   \\
\hspace{1mm}\\

$\Omega_k$ &  $-0.068\,_{-0.024}^{+0.058}$ &  $-0.013\,_{-0.007}^{+0.017}$  & $-0.0008\,^{+0.0038}_{-0.0050}$&  $-0.025\,_{-0.012}^{+0.015}$ &  $0.0005\pm 0.0064$ &  $-0.0056\, ^{+0.0063}_{-0.0075}$   \\
\hspace{1mm}\\

\hline
\hline

\end{tabular}}
\caption{$68 \% $ c.l.\ constraints on cosmological parameters in our extended $12$-parameter scenario, with spatial curvature $\Omega_k$ fit but $\alens=1$ fixed, from different combinations 
of data sets. If only upper limits are shown, they are at 95\% c.l. Note that $\sigma_8$ is a derived parameter.}
\label{table3}
\end{center}
\end{table*}

As seen in Table~\ref{table3} and as already discussed in \cite{hodev}, letting curvature freely vary brings the Hubble constant from Planck data alone to values that are incompatible with the R16 prior. 
This is essentially due to a well known geometrical degeneracy between the parameters $H_0$, $\Omega_k$, $w_0$ and $w_a$, but the same conclusion is obtained when combining the Planck data with BAO or JLA data sets.

\begin{figure*}
\centering
\begin{tabular}{cc}
\includegraphics[width=0.95\columnwidth]{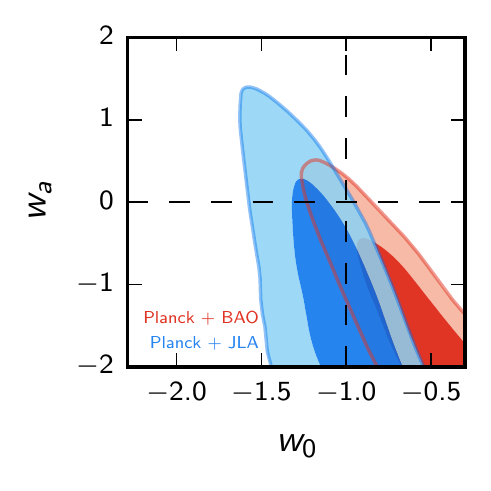} & \includegraphics[width=0.95\columnwidth]{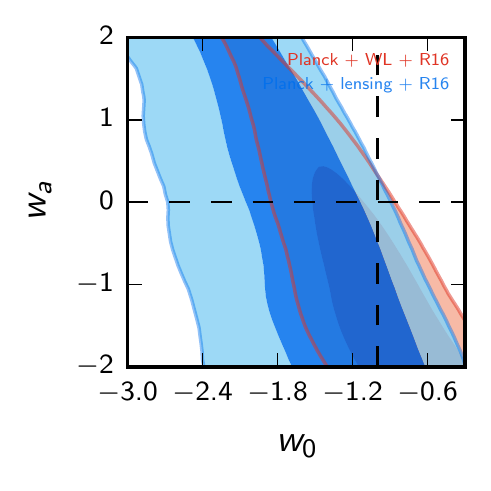}  \\
\end{tabular}
\caption{68.3\% and 95.4\% constraints on the $w_0$--$w_a$ plane in an $12$ parameter extended space, varying $\Omega_k$ but fixing $\alens=1$. 
The R16 prior is highly incompatible with the cosmological models preferred by the Planck+BAO and Planck+JLA datasets in this case. In the left panel,
we see that from these data sets   that the ($w_0>-1$, $w_a>0$) region is   barely excluded. In the right panel we show constraints from the Planck+R16+lensing and Planck+R16+WL data sets. These constraints are significantly weaker with respect to the previous cases, but the ($w_0>-1$, $w_a>0$) region is still excluded at nearly $95 \%$ c.l..
%The steep magenta line shows the "mirage" line giving the main geometric CMB degeneracy. The shallower orange half line shows the $w_0$-$w_a$ relation that many thawing quintessence models follow.
}
\label{fig3}
\end{figure*}

For dark energy, we see from Figure~\ref{fig3} that models with ($w_0>-1$, $w_a>0$) are disfavored by the Planck+BAO and Planck+JLA data sets even without the inclusion of the R16 prior. 
A similar conclusion is obtained when considering the Planck+R16+WL and Planck+R16+lensing data sets (see Figure~\ref{fig3}, right panel), even though the probability contours are significantly larger with respect to the previous cases considered.

We also note from the results in Table~\ref{table3} that there is no significant indication of any deviation from spatial flatness in all of the cases considered.  

\section{Conclusions}

We have  investigated the constraints on dynamical dark energy in an extended parameter space, considering the simultaneous variation of $12$ parameters. This is particularly of interest because in this extended parameter space the Planck and R16 datasets are consistent. Moreover, they point to a value for the dark energy equation-of-state $w<-1$. Note that in this extended parameter space, there is no preference for extra dark radiation, i.e.\ $N_{\rm eff}$ greater than the standard concordance value. 

Studying the dark energy equation-of-state phase space, we have indeed found that the Planck+R16 dataset not only rules out a cosmological constant at 95\% c.l.\ but also all standard quintessence  models, both freezing and the conventional thawing classes. This result is robust to different combinations of data, including the  WL, JLA, or lensing datasets.  Moreover, when the JLA dataset is included, also the remaining region  of "downwards-going" models ($w_a>0$) is disfavored at about $95 \%$ c.l.
A tension remains however with the BAO data set. The Planck+R16+BAO  dataset still allows a cosmological constant and a small portion of the freezing ($w_0>-1$, $w_a>0$) region. 

We have also tested the stability of these results through two further variations. 
Restricting to a smaller $11$-parameter space by fixing the lensing amplitude $\alens=1$, the results hold with just a reduction 
of the available model volume. In  this case also, the freezing ($w_0>-1$, $w_a>0$) region starts to be incompatible with the Planck+R16+BAO data set. Secondly, we then allowed 
spatial curvature as a free parameter. When $\Omega_k$ is varied, the Planck+BAO and Planck+JLA datasets provide values for the Hubble constant that are no longer compatible with the R16 prior. We found that also in this case when considering the constraints on $w(a)$, the freezing region is not favored by the data, though the thawing region is only mildly disfavored.
The same conclusion in the $12$-parameter space that includes curvature is obtained from the Planck+R16+lensing and Planck+R16+WL datasets.  

In summary, taking all data sets at face value, we find that both the freezing class of quintessence  and the region of parameter space typical of the thawing class of quintessence are generally disfavored. One needs either $w_0>-1$ but highly negative $w_a$ (as preferred, say, by Planck+JLA+R16) or $w_0<-1$. Even the first option will also have $w<-1$ at some redshift, so phantom models of dark energy seem preferred. In Di Valentino et al.\ 2017b (in preparation), we consider some physically motivated models that can match the results of the current Planck+R16 data set, in particular particle physics phase transition models. 

Conversely, we have shown how the preferred region of $w_0$--$w_a$ phase space shifts for various values of the Hubble constant, if further measurements or reanalyses change the current prior. In particular, for $H_0<70\,$km/s/Mpc, the cosmological constant lies within the 68\% c.l., and regions of standard freezing and thawing quintessence are acceptable as well. 

Before concluding, it is worth mentioning that small, unresolved systematics can be easily present in all the datasets we have considered. 
The R16 estimate of the Hubble constant is based on the combination of three different geometric distance calibrations of Cepheids \cite{R16}. These three different methods yield three constraints on the Hubble constant: $H_0=72.25\pm 2.51$ km/s/Mpc, $H_0=72.04\pm2.67$ km/s/Mpc, and $H_0=76.18\pm2.37$ km/s/Mpc (again, see \cite{R16}).  
Discarding the last constraint (based on Milky Way Cepheids) could reduce the current tension and, therefore, change our conclusions (as we showed with Figure~\ref{figH0}). While there is currently no reason to remove the Milky Way constraint, this emphasizes that the results reported here can be driven just by one portion of the R16 data. A similar result occurs with BAO: taking the four BAO datasets separately, we have found that while the 6dFGS, SDSS-MGS, and BOSS-LOWZ sets are in agreement with the Planck+R16 solution, the major discrepancy comes from the CMASS-DR11 single data point (and even more so for DR12 \cite{alam}). See, e.g., \cite{hirata,beutler} for recent discussion of possible BAO systematics. 
On the other hand, the nature of the Planck $\alens$ anomaly is still unclear. The Planck lensing dataset is in tension with the Planck angular power spectra data and this tension persists also in our 12 parameters analysis. While it does not appear to strongly affect our bounds on the equation of state it may not be optimally described by a single parameter (see Di Valentino et a.\ 2017b, in preparation). Thus, there is still much to learn about both the nature of dark energy and the robustness of data sets.

\section{Acknowledgments}
This work has been done in part within the Labex ILP (reference ANR-10-LABX-63) part of the Idex SUPER, and received financial state aid managed by the Agence Nationale de la Recherche, as part of the programme Investissements d'avenir under the reference ANR-11-IDEX-0004-02. The work of EDV and J.S. has been supported in part by ERC Project
No. 267117 (DARK) hosted by the Pierre and Marie Curie
University-Paris VI, Sorbonne Universities and CEA-Saclay and by the Institut Lagrange de Paris. 
EL was supported in part by the Energetic Cosmos Laboratory and by 
the U.S.\ Department of Energy, Office of Science, Office of High Energy 
Physics, under Award DE-SC-0007867 and contract no.\ DE-AC02-05CH11231. 

%\eric{NOTE: Refs.\ 10 and 24-29 do not appear in the text. Should we remove them?}

\end{document}